\documentclass[%
 aip,
 jmp,%
 reprint,%
]{revtex4-2}

\usepackage{graphicx}
\usepackage{dcolumn}
\usepackage{bm}
\usepackage{color}
\usepackage{hyperref}

\begin{document}

\title{Enhanced plasma current spike formation due to onset of $1/1$ kink-tearing reconnection during a massive gas injection process}

\author{Shiyong Zeng}
\affiliation{Department of Engineering and Applied Physics, University of Science and Technology of China, Hefei, Anhui 230026, China}

\author{Ping Zhu}
\email{zhup@hust.edu.cn}
\affiliation{International Joint Research Laboratory of Magnetic Confinement Fusion and Plasma Physics, State Key Laboratory of Advanced Electromagnetic Engineering and Technology, School of Electrical and Electronic Engineering, Huazhong University of Science and Technology, Wuhan, Hubei 430074, China}
\affiliation{Department of Engineering Physics, University of Wisconsin-Madison, Madison, Wisconsin 53706, USA}

\author{Haijun Ren}
\email{hjren@ustc.edu.cn}
\affiliation{Department of Engineering and Applied Physics, University of Science and Technology of China, Hefei, Anhui 230026, China}

\date{\today}

\begin{abstract}
	The formation of the plasma current spike at the end of the thermal quench phase is studied systematically, which is found to strongly correlate to the onset of the $1/1$ kink-tearing reconnection in the simulation results. The magnetohydrodynamic (MHD) activity on the $q = 1$ surface plays a critical role in the spike formation and the disruption process, namely, when the safety factor in the magnetic axis $q_0$ exceeds 1, the plasma major disruption transits into successive minor disruptions and the start of thermal quench phase is delayed.
\end{abstract}

\maketitle

\section{\label{sec:introduction}Introduction}
It is well known that the plasma disruption in tokamak would cause irreversible destruction to the device, e.g. the consequential thermal load, electromagnetic force, and the extremely high energetic runaway electrons (RE) to destroy the plasma facing component (PFC). Mitigation strategies in case of inevitable disruptions events have been developed over decades, such as the massive gas injection (MGI) \cite{Gerasimov_2020} and the shattered or shell Pellet Injection (SPI or ShPI) \cite{Hollmann2019,Commaux_2016}, among others. The formation of a plasma current spike is universally observed at the end of the thermal quench (TQ) stage and the beginning of the current quench (CQ) stage during the disruption process, which means a large loop voltage is induced, and such a spike is considered to correlate with the generation of the RE population \cite{Svenningsson2021,Decker_2022}. Thus, the origin and the mechanism of the current spike formation, as well as its key parameter dependence, has been a subject of continued interests.

The $q$ profile dependence in the current spike is observed in D\uppercase\expandafter{\romannumeral3}-D experiments, which shows the amplitude of the current spike increases with the reduced depth of the $q=2$ surface from the separatrix \cite{Hollmann2007}. A model is used to explain this spike in terms of an initial trapping of poloidal flux and its subsequent release due to a sudden edge cooling of the plasma \cite{Wesson_1990}. Latter, JOREK simulations show that the current relaxation in the region $q \le 2$ along with the growing $m=2/n=1$ (here $m/n$ is the poloidal/toroidal mode number) magnetic island generates a plasma current spike \cite{Nardon2021}. Previous NIMROD simulations demonstrate the correlation between the formation of the current spike and the magnetic reconnection activity in the core region \cite{Izzo_2010,Izzo2021}, and we further identify the onset of a $1/1$ kink-tearing instability as an underlying internal reconnection process in close association to the current spike formation in this study.

In particular, the formation of the current spike at the end of the thermal quench phase is reproduced using the NIMROD code in this work, and the effects of $q_0$ on the current spike formation are studied. When the $q_0 <1$, a $1/1$ kink-tearing mode is localized in the central region and one major disruption occurs at the end of thermal quench stage, and the amplitude of the current spike increases with the decreased value of $q_0$. When the $q_0$ exceeds 1, a quasi-interchange like mode arises in the core region and the plasma undergoes two successive minor disruptions at the end of thermal quench phase.

The rest of this paper is arranged as follows, section \ref{sec:simulation model} introduces the simulation model and setup, section \ref{sec:resluts} shows the simulation results and demonstrates the role of $q_0$ on the formation of the current spike and the disruption process, and section \ref{sec:summary} gives a summary.

\section{\label{sec:simulation model}Simulation model and setup}
Our simulations are based on the 3D resistive MHD model implemented in the NIMROD code \cite{SOVINEC2004} along with the impurity radiation module KPRAD \cite{KPRAD}, and the combined system of equations can be written as follows \cite{Izzo2008}
\begin{eqnarray}
	\rho \frac{d \vec{V}}{dt} = - \nabla p + \vec{J} \times \vec{B} + \nabla \cdot (\rho \nu \nabla \vec{V})
	\label{eq:momentum}
	\\
	\frac{d n_i}{d t} + n_i \nabla \cdot \vec{V} = \nabla \cdot (D \nabla n_i) + S_{ion/3-body}
	\label{eq:contiune2}
	\\
	\frac{d n_{Z}}{d t} + n_Z \nabla \cdot \vec{V} = \nabla \cdot (D \nabla n_Z) + S_{ion/rec}
	\label{eq:contiune3}
	\\
	n_e \frac{d T_e}{d t} = (\gamma - 1)[n_e T_e \nabla \cdot \vec{V} + \nabla \cdot \vec{q_e} - Q_{loss}]
	\label{eq:temperature}
	\\
	\vec{q}_e = -n_e[\kappa_{\parallel} \hat{b} \hat{b} + \kappa_{\perp} (\mathcal{I} - \hat{b} \hat{b})] \cdot \nabla T_e
	\label{eq:heat_flux}
	\\
	\frac{\partial \vec{B}}{\partial t} = \nabla \times \left( \vec{V} \times \vec{B} \right) - \nabla \times \left(  \eta \vec{j} \right) 
	\label{eq:ohm}
\end{eqnarray}	
Here, $n_i$, $n_e$, and $n_Z$ are the main ion, electron, and impurity ion number density respectively, $\rho$, $\vec{V}$, $\vec{J}$, and $p$ the plasma mass density, velocity, current density, and pressure respectively, $T_e$ and $\vec{q}_e$ the electron temperature and heat flux respectively, $D$, $\nu$, $\eta$, and $\kappa_{\parallel} (\kappa_{\perp})$ the plasma diffusivity, kinematic viscosity, resistivity, and parallel (perpendicular) thermal conductivity respectively, $\gamma$ the adiabatic index, $S_{ion/rec}$ the density source from ionization and recombination, $S_{ion/3-body}$ also includes contribution from 3-body recombination, $Q_{loss}$ the energy loss, $\vec{E} (\vec{B})$ the electric (magnetic) field, $\hat{b}=\vec{B}/B$, and $\mathcal{I}$ the unit dyadic tensor.

\begin{table}[b]
	\caption{\label{tab:input} Key parameters in the simulation}
	\begin{ruledtabular}
		\begin{tabular}{cccccccc}
			\textbf{Parameter} & \textbf{Symbol} & \textbf{Value} & \textbf{Unit} \\
			\hline
			Minor radius & $a$ & $0.25$ & $m$ \\
			Major radius & $R_0$ & $1.05$ & $m$ \\
			Plasma current & $I_p$ & $150$ & kA \\
			Toroidal magnetic field & $B_{t0}$ & $1.75$ & T\\
			Edge value of safety factor & $q_a$ & $3.56$ & dimensionless \\
			Core electron density & $n_{e0}$ & $1.875 \times 10^{19}$ & $m^{-3}$ \\
			Core electron temperature & $T_{e0}$ & $700$ & $eV$\\
			Equilibrium velocity & $V_0$ & $0$ & $m/s$\\
			The core resistivity & $\eta_0$ & $5.1285 \times 10^{-9}$  & $\Omega \cdot m$  \\
			Kinematic viscosity & $\nu$ & $27$  & $m^2/s$  \\
			The core Lundquist number & $S_0$ & $3.8468 \times 10^8$  & dimensionless \\
			The core Prandtl number & $P_r$ & $6.6158 \times 10^3$  &dimensionless  \\
			Core perpendicular thermal conductivity & $\kappa_{\perp0}$ & $1$ & $m^2/s$ \\
			Core parallel thermal conductivity & $\kappa_{\parallel0}$ & $10^6$ & $m^2/s$ \\
			Diffusivity & $D$ & $2$ & $m^2/s$\\
		\end{tabular}
	\end{ruledtabular}
\end{table}

\begin{figure}[ht]
	\begin{center}
		\includegraphics[width=1.0\textwidth,height=0.55\textheight]{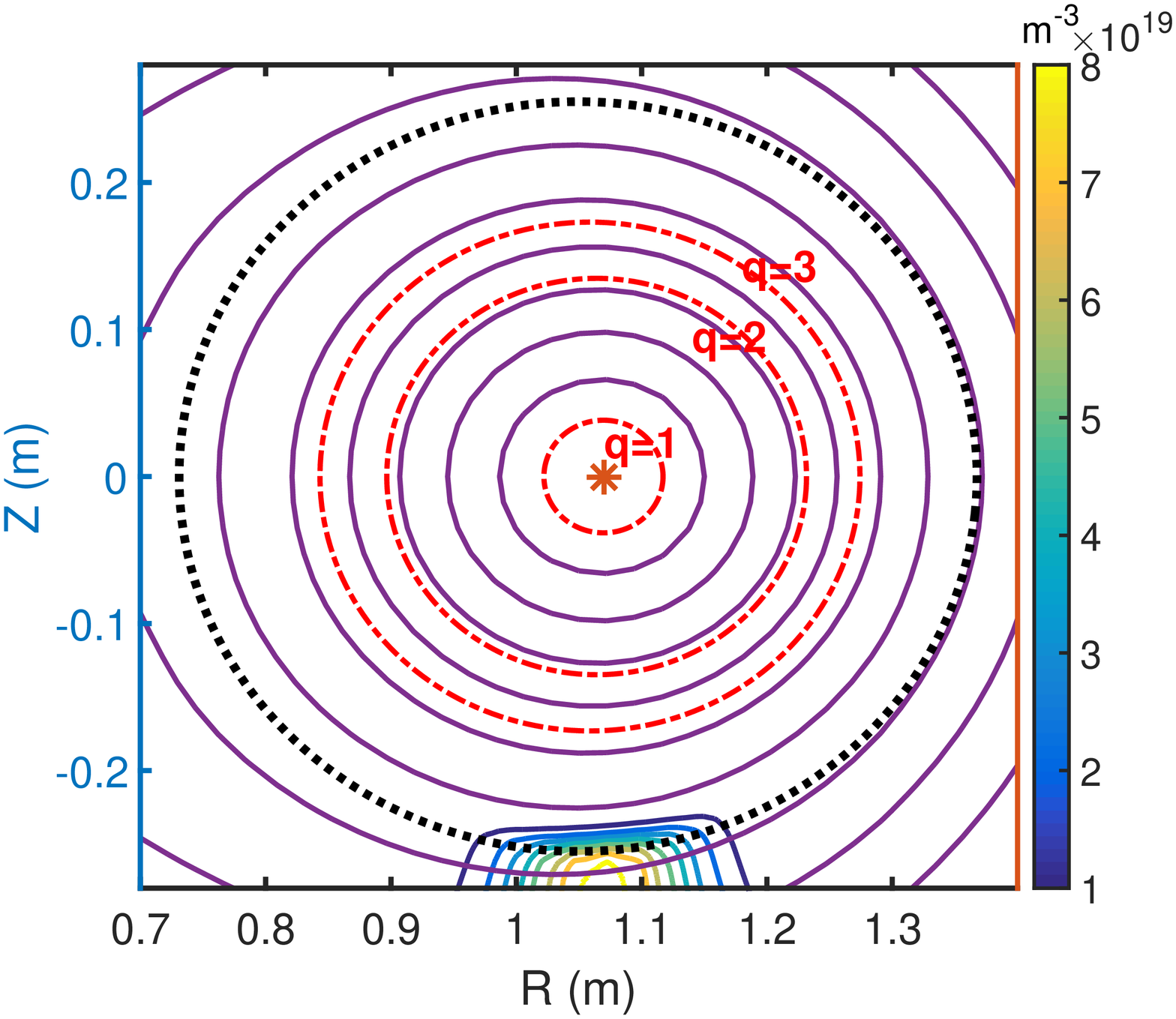}
	\end{center}
	\caption{Equilibrium magnetic flux $\psi$ contour (purple solid lines) and impurity density distribution (in unit $m^{-3}$, flushed color), the equilibrium $q = 3, 2, 1$ surfaces are denoted as red dashed lines and the boundary of simulation domain is denoted as black dotted line.}
	\label{fig:psi}
\end{figure}

The equilibrium is based on a circular-shaped limiter tokamak plasma and some key parameters are listed in table. \ref{tab:input}. A massive gas injection triggered disruption process is simulated and the impurity is injected from the bottom (Fig. \ref{fig:psi}). The neutral source is static and localized both along the toroidal, poloidal and radial directions at the beginning of the simulation. Then, the ionization and recombination of the neutral occur with the inward penetration of the impurity, which mainly depends on the diffusion and convection. The Spitzer resistivity $\eta \propto T_e^{-3/2}$ and the Braginskii scaling of thermal conductivity $\kappa_{\perp} \propto T_e^{-1/2}B^{-2}$, and $\kappa_{\parallel} \propto T_e^{5/2}$ are adopted.
A set of simulations with different value of $q_0$ are performed, and the entire equilibrium $q$ profile shifts up and down with the change of the $q_0$ only and its shape fixed.
We use $64 \times 63$ bicubic Lagrange polynomial finite elements in the poloidal plane, and six Fourier modes with toroidal numbers $n=0-5$ in the toroidal direction. The plasma is limited by a perfect conducting wall, and the boundary of the simulation domain is surrounded by a vacuum region.

\section{\label{sec:resluts}Simulation results}

\subsection{\label{subsec:history}Time history of massive gas injection process}
\begin{figure}[ht]
	\begin{center}
		\includegraphics[width=0.9\textwidth,height=0.8\textheight]{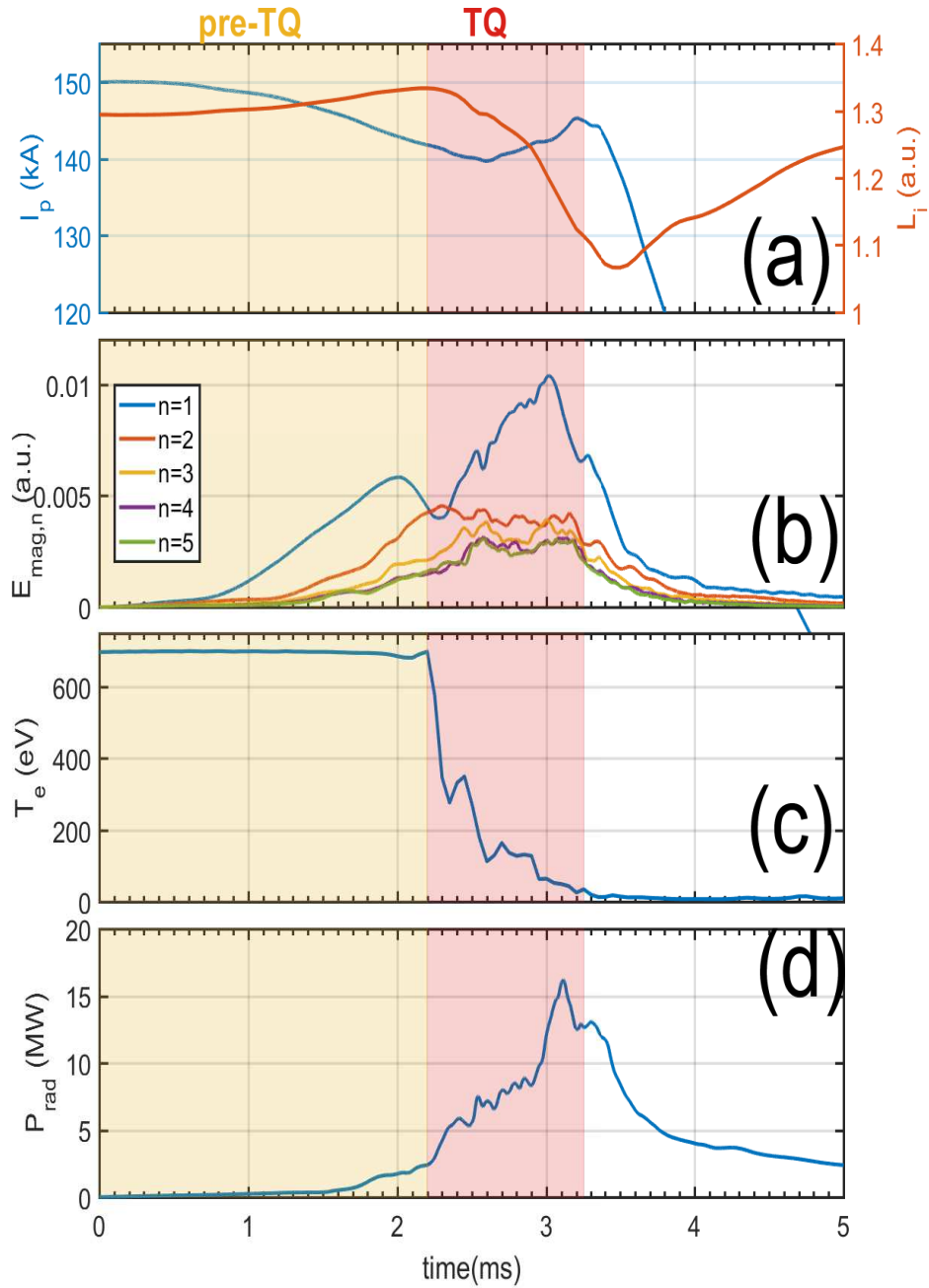}
	\end{center}
	\caption{(a) Plasma current (blue solid line) and internal inductance (red solid line), (b) normalized magnetic energies of toroidal components $E_{mag,n} = \sqrt(W_{mag,n}/W_{mag,n=0})$, (c) core electron temperature, and (d) radiation power as functions of time during a MGI process from NIMROD simulation, where $0-2.2 ms$ is the pre-TQ phase (yellow shade), and $2.2-3.25 ms$ is the TQ phase (red shade).}
	\label{fig:discharge}
\end{figure}

Representative MGI mitigated disruption is shown in Fig. \ref{fig:discharge}, which reproduces the main features of this dynamic process throughout the thermal quench (TQ) phase. During the pre-thermal quench (pre-TQ) phase, the impurity radiation power and the perturbed magnetic energy increases gradually along with impurity inward penetration, and the $n=1$ mode dominates. Then, the sudden collapse of the central electron temperature at $t=2.2ms$ leads to the initiation of the TQ stage, and the radiation power and the perturbed magnetic energy reach their maximum values respectively right after $t=3ms$. The subsequent appearance of the plasma current spike around $t=3.25ms$ typically signifies the end of TQ stage and the beginning of the CQ phase.

\subsection{\label{subsec:mechannism}Formation of current spike}
\begin{figure}[ht]
	\begin{center}
		\includegraphics[width=1.0\textwidth,height=0.35\textheight]{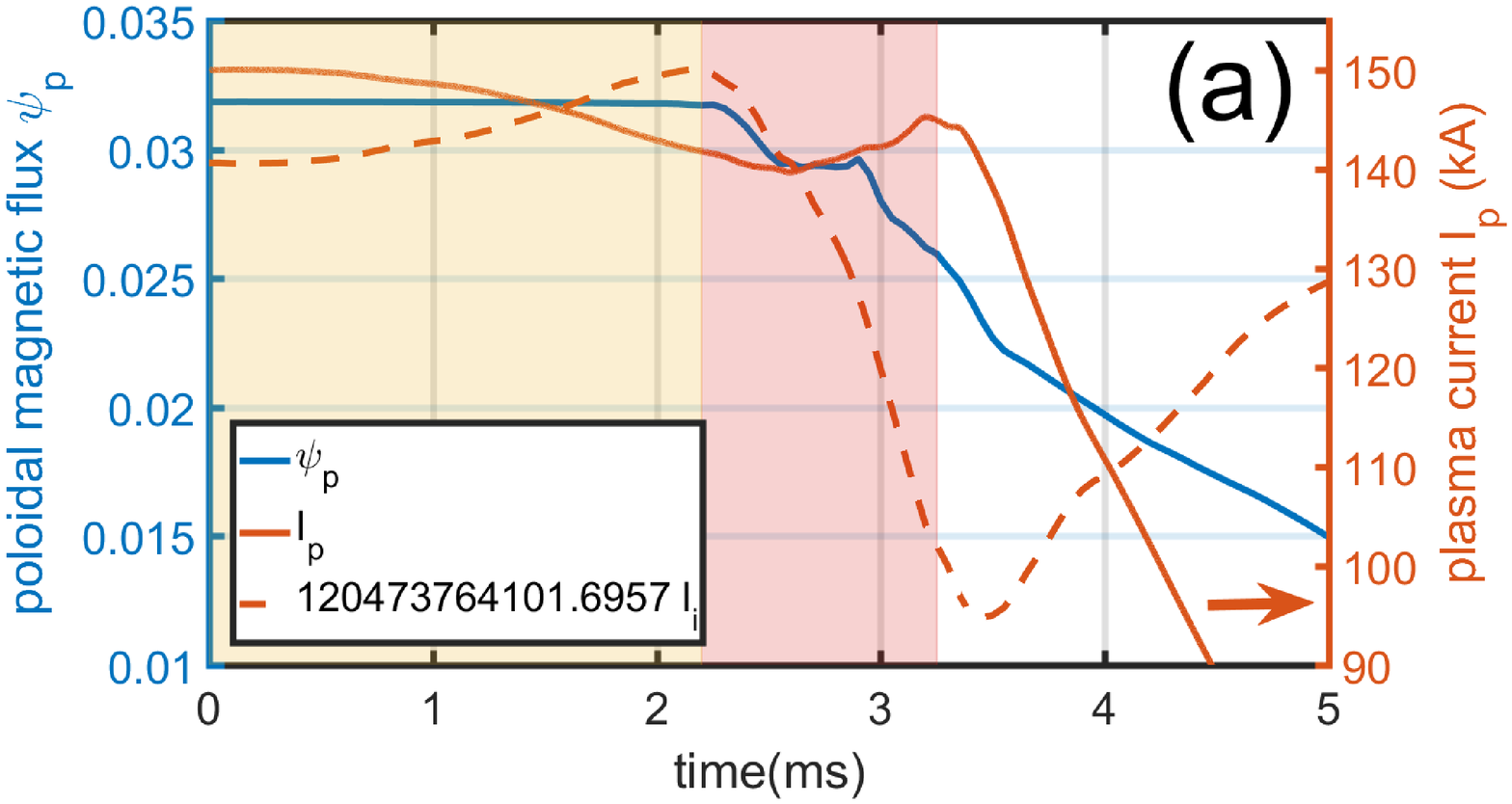}
		\includegraphics[width=1.0\textwidth,height=0.35\textheight]{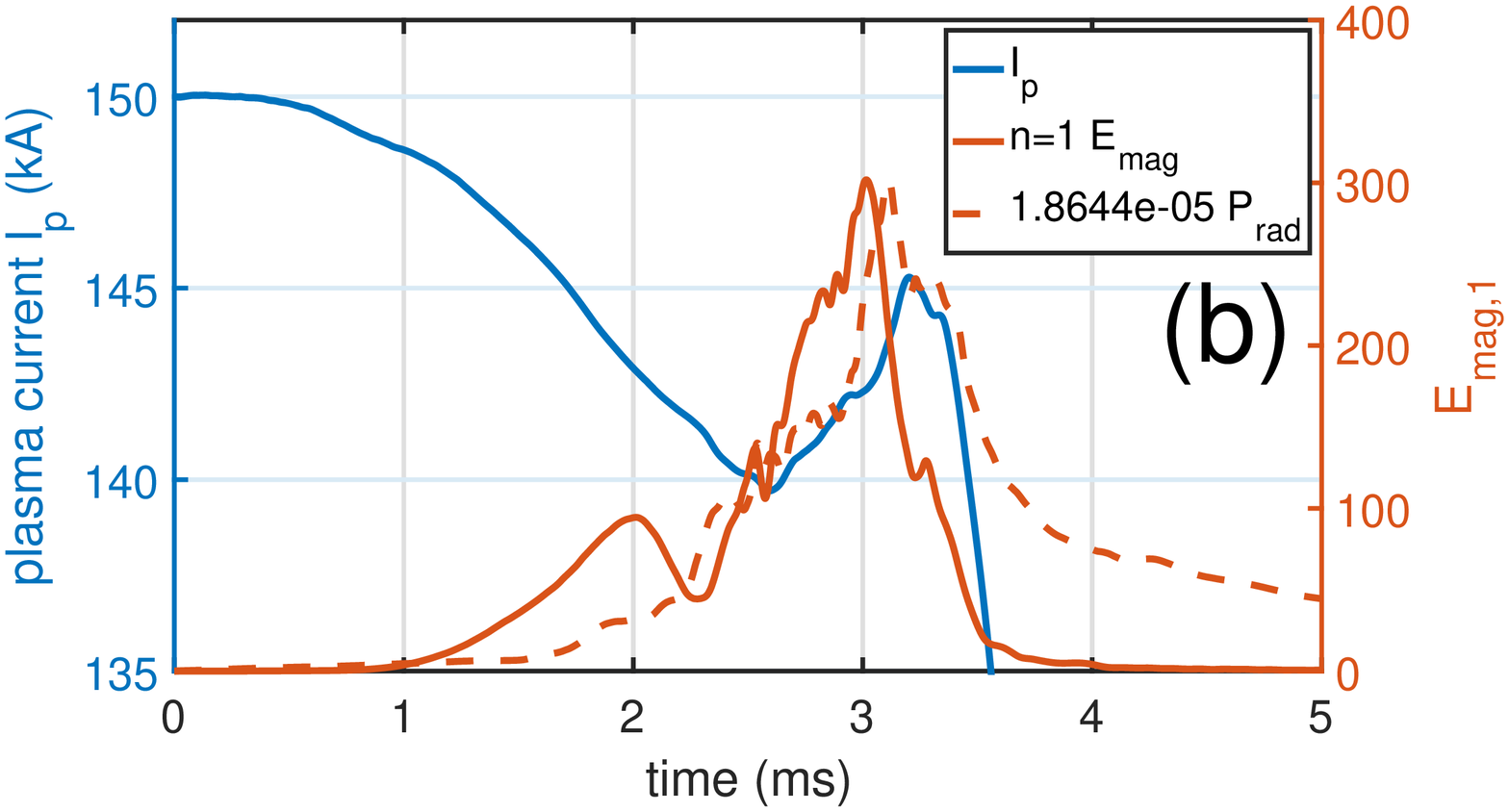}
	\end{center}
	\caption{(a) Poloidal magnetic flux (blue solid line), Plasma current (orange solid line), and internal inductance (orange dashed line), and (b) plasma current (blue solid line), perturbed magnetic energies of toroidal component $n=1$ (orange solid line), and impurity radiation power (orange dashed line) as functions of time.}
	\label{fig:plasma current}
\end{figure}

Generally, the current spike formation is attributed to the magnetic flux conservation $\Psi_p = L_i I_p$, where the plasma current $I_p$ and plasma internal inductance $L_i=\mu_0R_0\int_{0}^{2\pi}\int_{0}^{a} B_{\theta}^2 rdrd\theta/(2\pi a^2B_{\theta a}^2)$, $\mu_0$ is the permeability of vacuum, $B_{\theta}$ is the poloidal magnetic field and $B_{\theta a}$ is its value at plasma edge $r=a$. As shown in Fig. \ref{fig:plasma current}(a), the internal inductance $L_i$ increases as result of the current profile contraction due to radiation cooling during the pre-TQ phase, and the plasma current $I_p$ decreases slightly. Then, the internal inductance $L_i$ drops suddenly as the current profile expands outwards at the TQ stage, and the current spike arises simultaneously. We note that the flux $\Psi_p$ is well conserved during the pre-TQ phase with higher temperature, but starts to decay in the TQ stage, as a result of the enhanced resistive diffusion brought by the electron temperature collapse and the magnetic reconnection in the central core region. From Fig. \ref{fig:plasma current}(b), the amplitude of the $n=1$ magnetic perturbation energy grows to its maximum right before the current spike, which suggests the correlation between the central MHD activity and the formation of the current spike.

\subsection{\label{subsec:q0_scaling}Role of $q_0$}
\begin{figure}[ht]
	\begin{center}
		\includegraphics[width=0.85\textwidth,height=0.45\textheight]{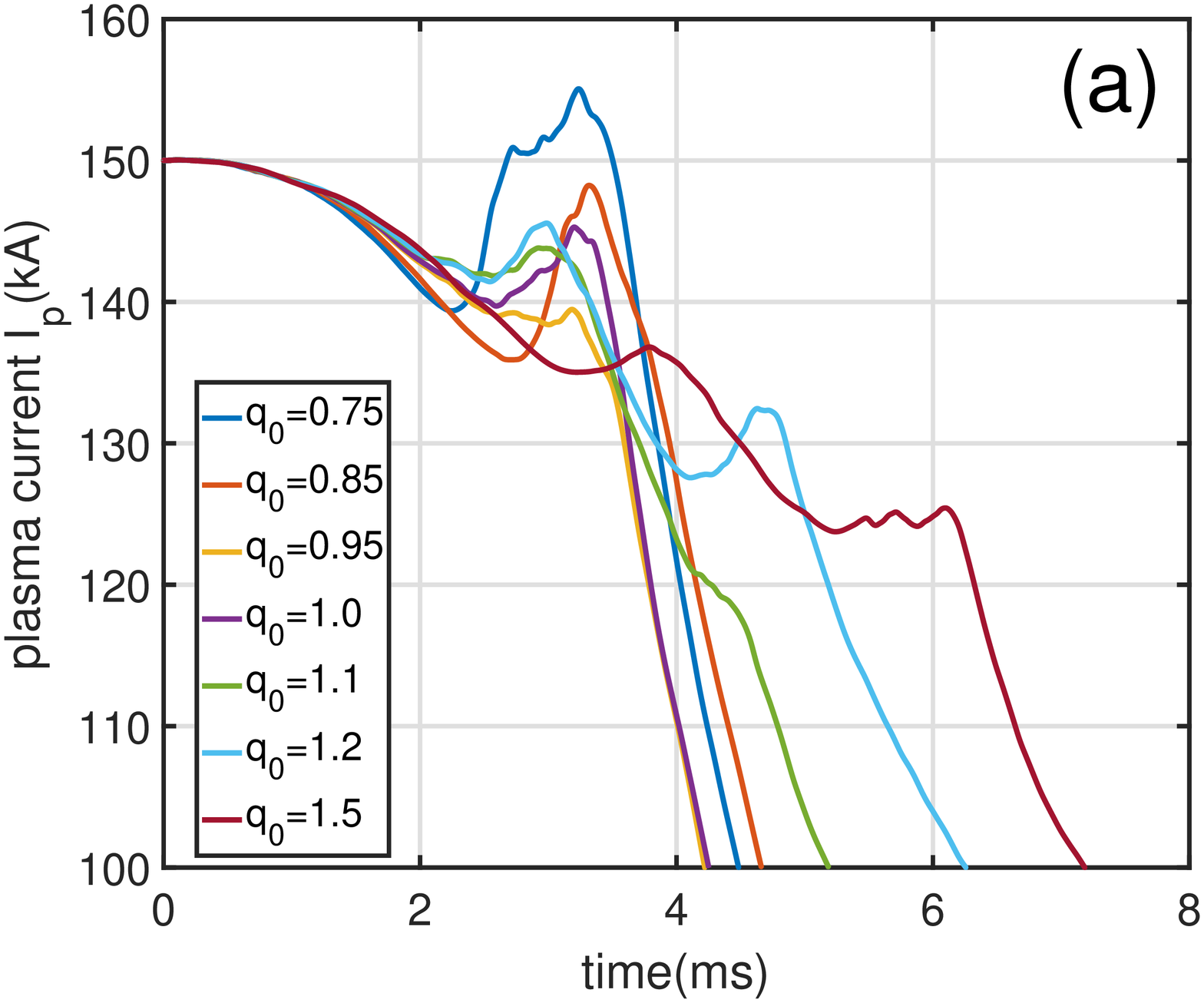}
		\includegraphics[width=0.85\textwidth,height=0.45\textheight]{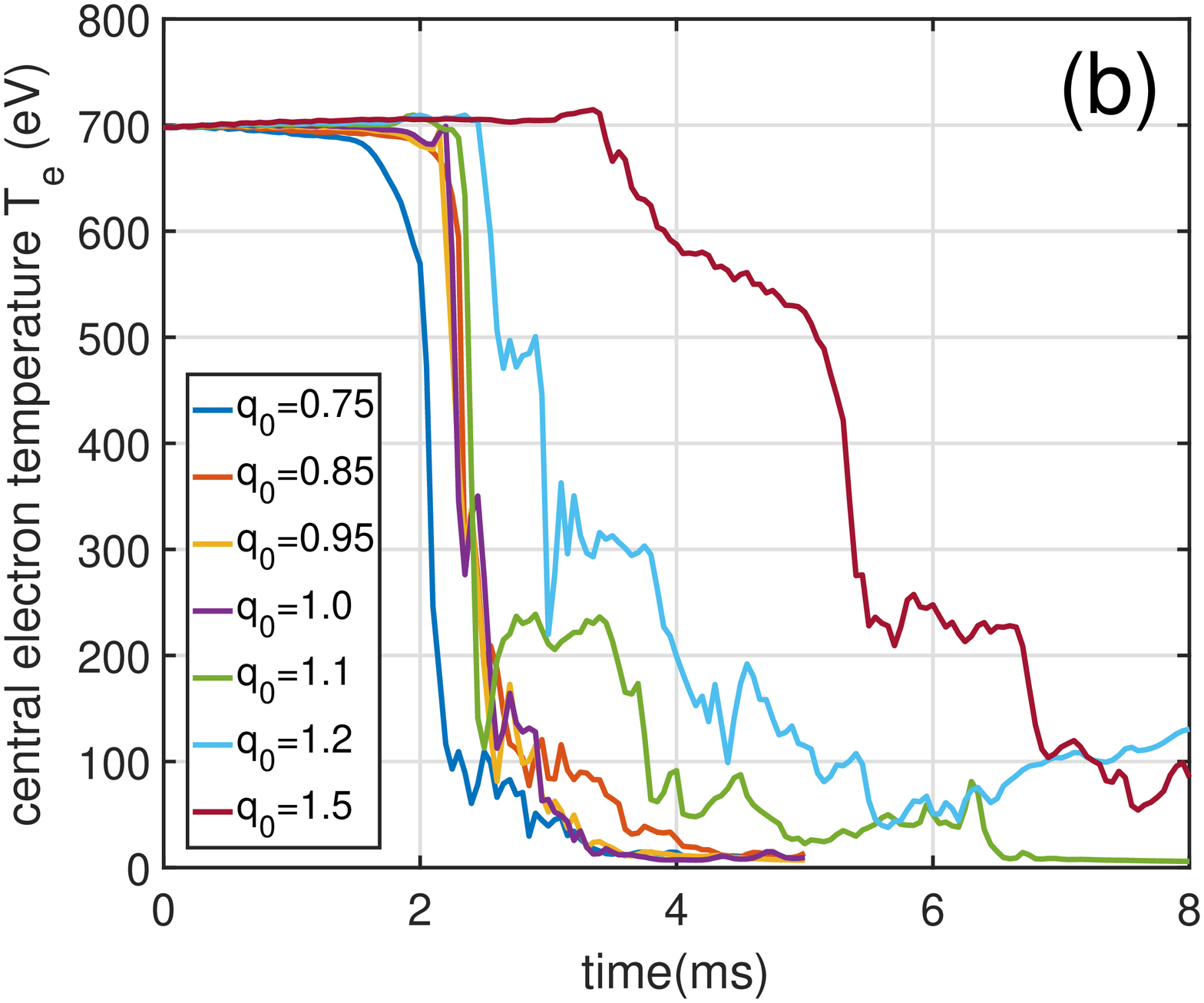}
	\end{center}
	\caption{(a) Plasma current, and (b) central electron temperature as functions of time for cases with different value of $q_0$.}
	\label{fig:q0-Ip-Te}
\end{figure}

\begin{figure}[ht]
	\begin{center}
		\includegraphics[width=0.45\textwidth,height=0.3\textheight]{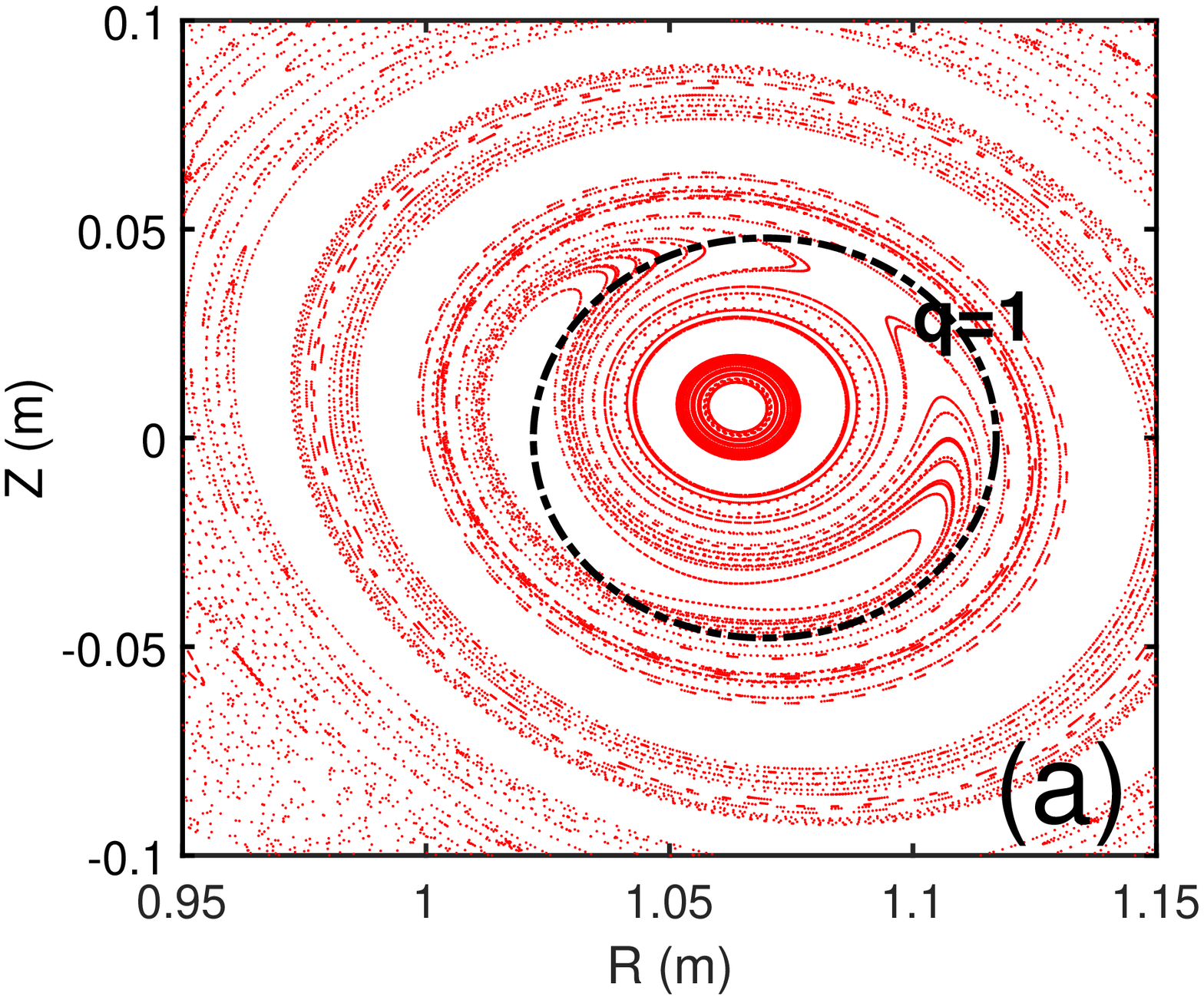}
		\includegraphics[width=0.45\textwidth,height=0.3\textheight]{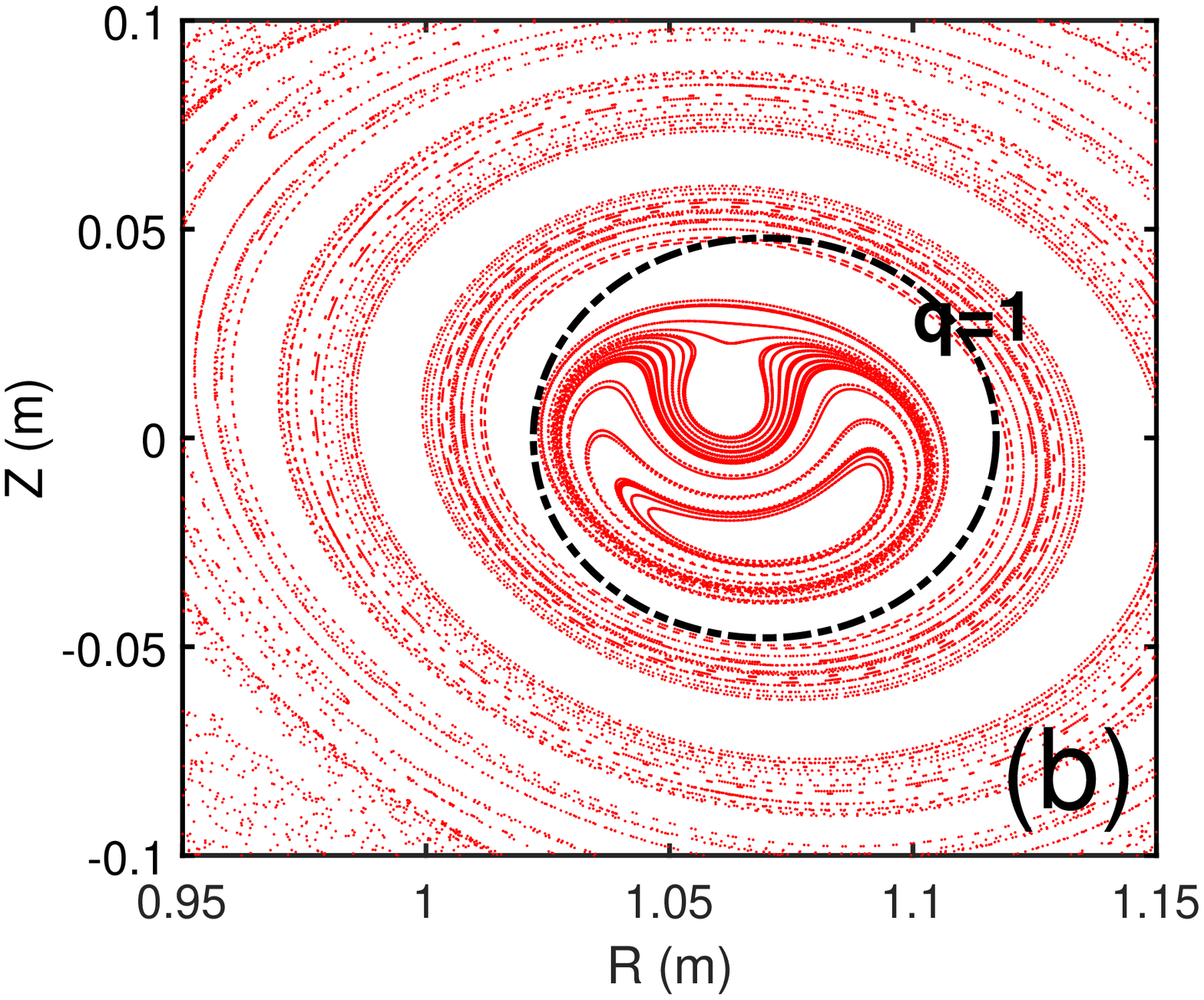}
		\includegraphics[width=0.45\textwidth,height=0.3\textheight]{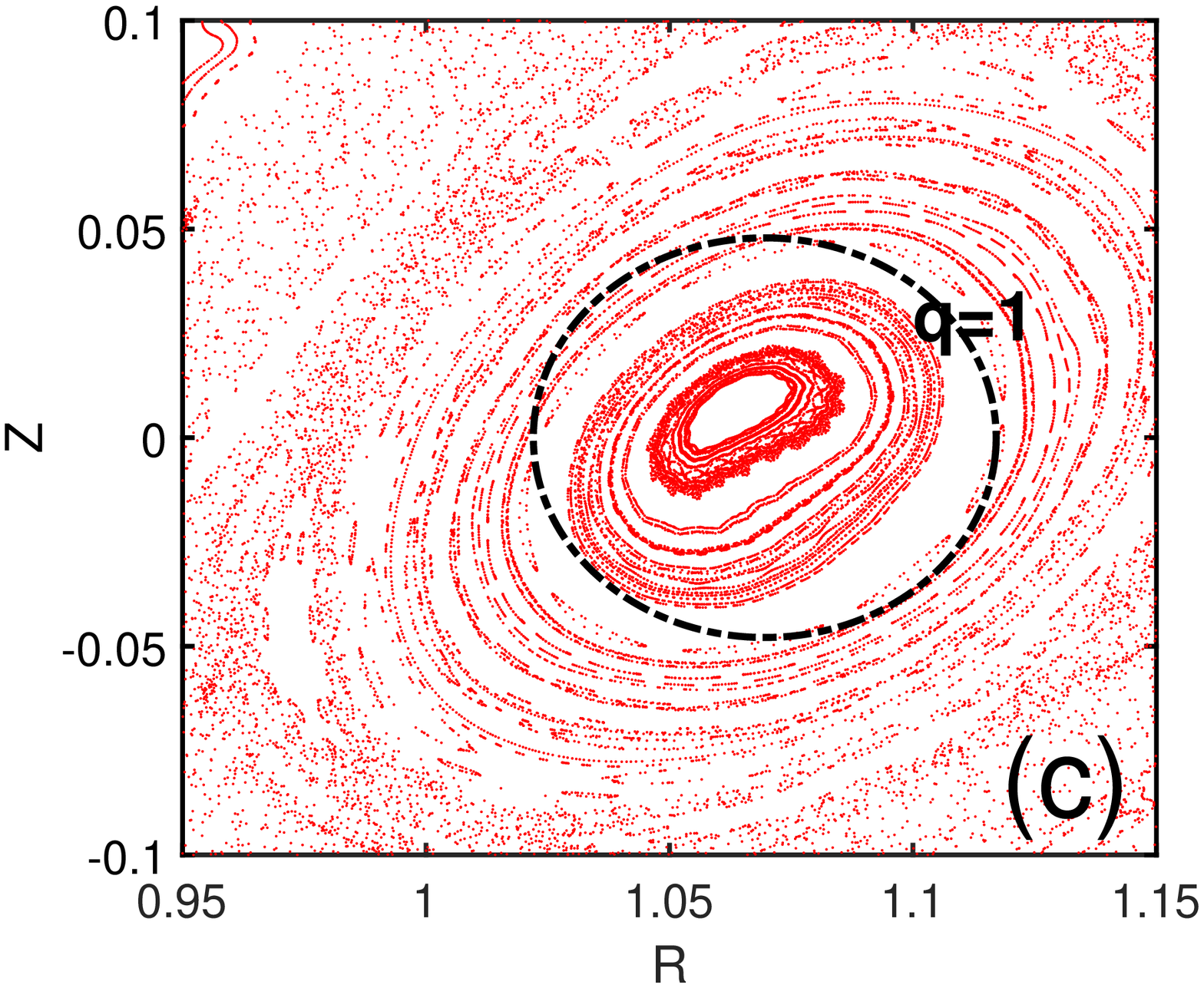}
	\end{center}
	\caption{The Poincare plots in the central region of cases with (a) $q_0=0.95$, (b) $q_0=1.0$, and $q_0=1.1$. The equilibrium $q=1$ surface of the case with $q_0=0.95$ is imposed as black dashed circle for the frame of reference in all plots.}
	\label{fig:He-q0-poincare}
\end{figure}

\begin{figure}[ht]
	\begin{center}
		\includegraphics[width=0.85\textwidth,height=0.45\textheight]{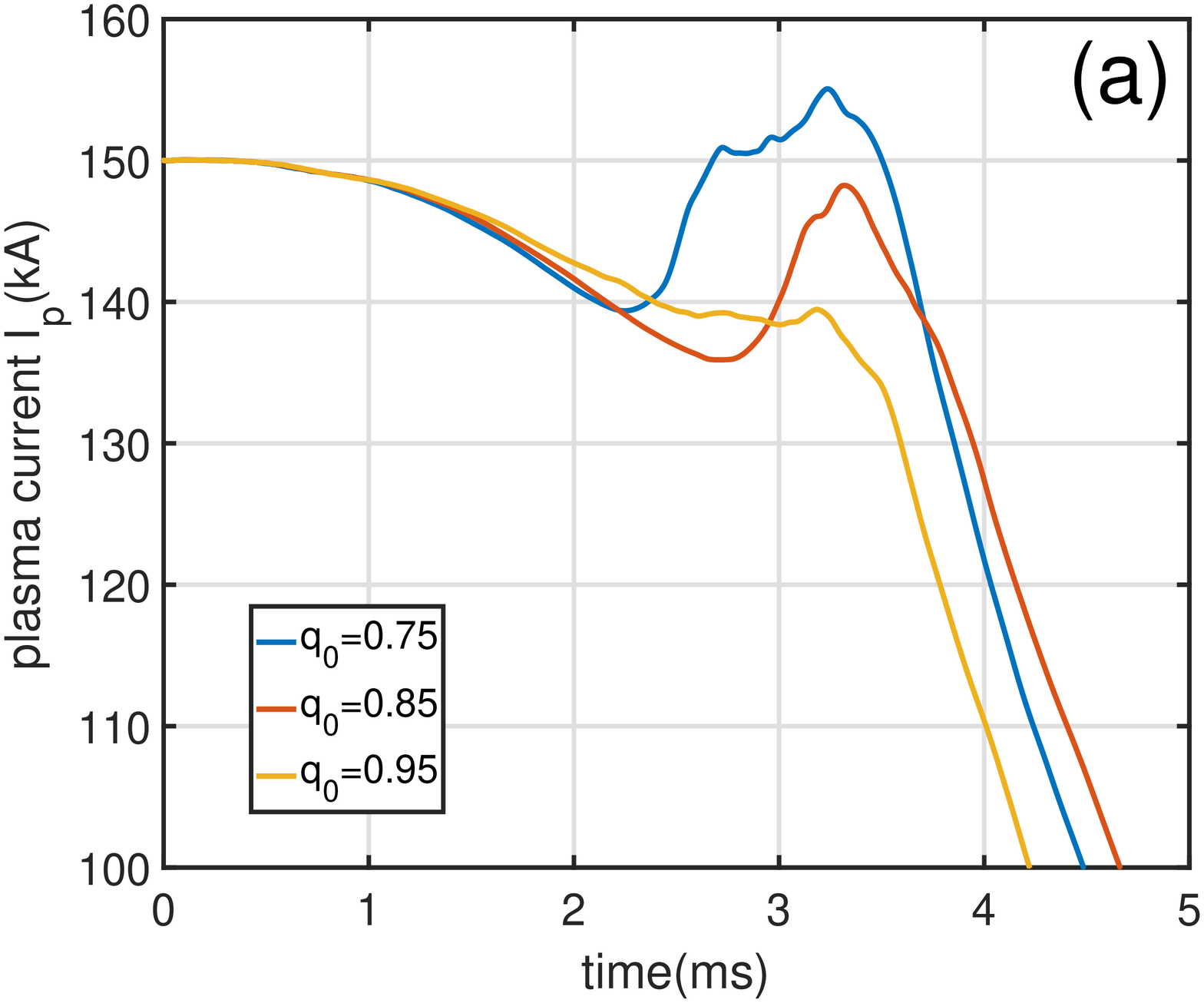}
		\includegraphics[width=0.85\textwidth,height=0.45\textheight]{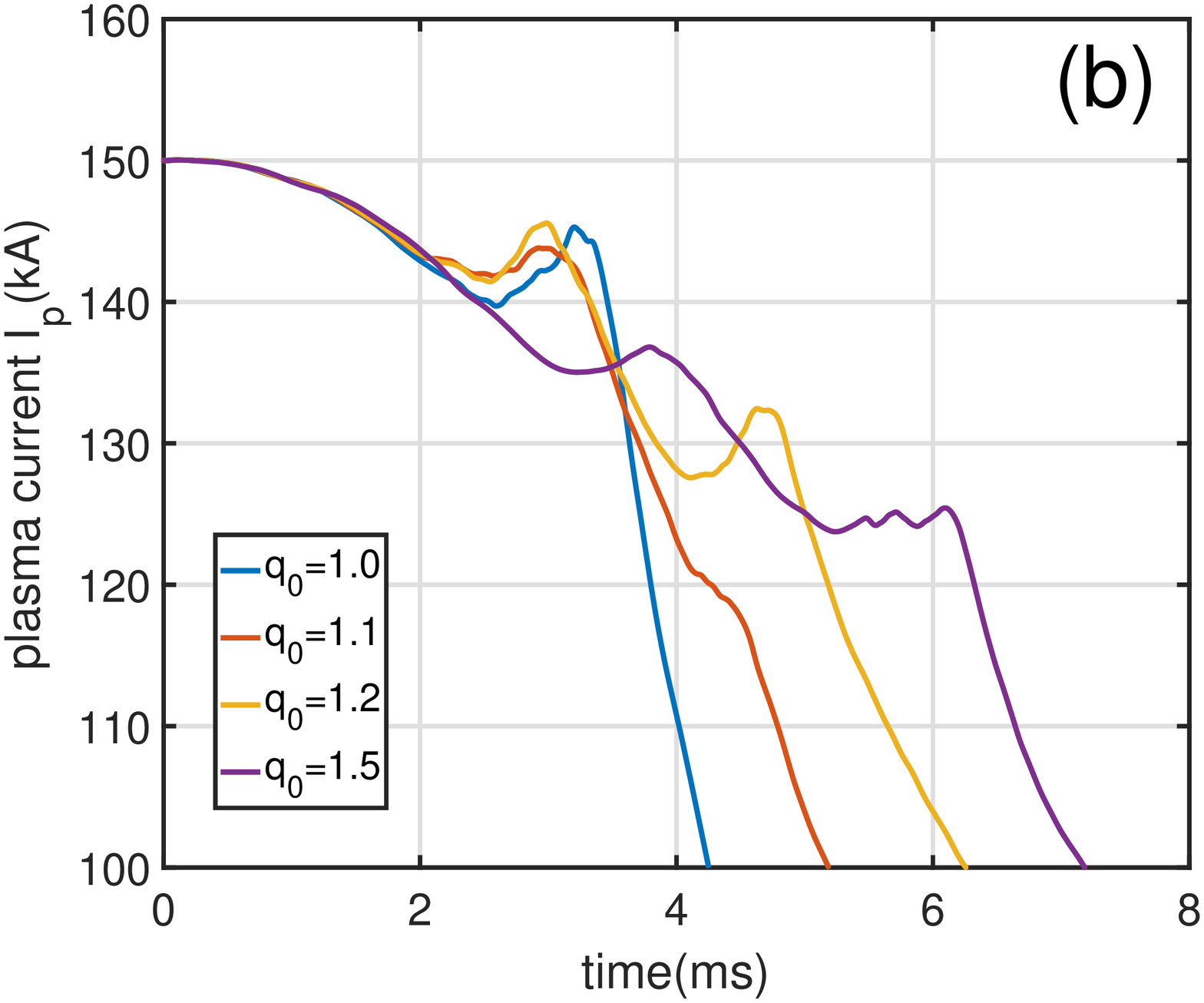}
	\end{center}
	\caption{The plasma current as functions of time for cases with (a) $q_0<1$, and (b) $q_0 \ge 1$.}
	\label{fig:samll-large-q0}
\end{figure}

The amplitude of the current spike depends on the value of $q_0$ (Fig. \ref{fig:q0-Ip-Te}). More importantly, only one spike forms when $q_0 \le 1$, and the major disruption occurs, whereas two or more spikes arise when $q_0>1$, which is followed by the minor disruption. As indicated by the central temperature collapse as well, the temperature drops through oscillations before the complete cooling in the case when $q_0$ exceeds 1 (Fig. \ref{fig:q0-Ip-Te}b).
From the poincare plot (Fig. \ref{fig:He-q0-poincare}), the transition from major disruption to minor disruption is shown to correlate to the central MHD activity. When the $q_0 \le 1$, a kink-tearing instability is localized in the central region, while in the cases with $q_0 > 1$, it turns into quasi-interchange in absence of the formation of the $(1,1)$ island structure.
In addition, the amplitude of the current spike increases as $q_0$ drops lower below unity (Fig. \ref{fig:samll-large-q0}a), which is consistent with the experimental observations \cite{Hollmann2007}. The time duration of the TQ stage becomes longer when $q_0$ increases above one (Fig. \ref{fig:samll-large-q0}b), suggesting the plasma is more stable when the $q=2$ surface locates further towards the magnetic axis.

\begin{figure}[ht]
	\begin{center}
		\includegraphics[width=1.0\textwidth,height=0.55\textheight]{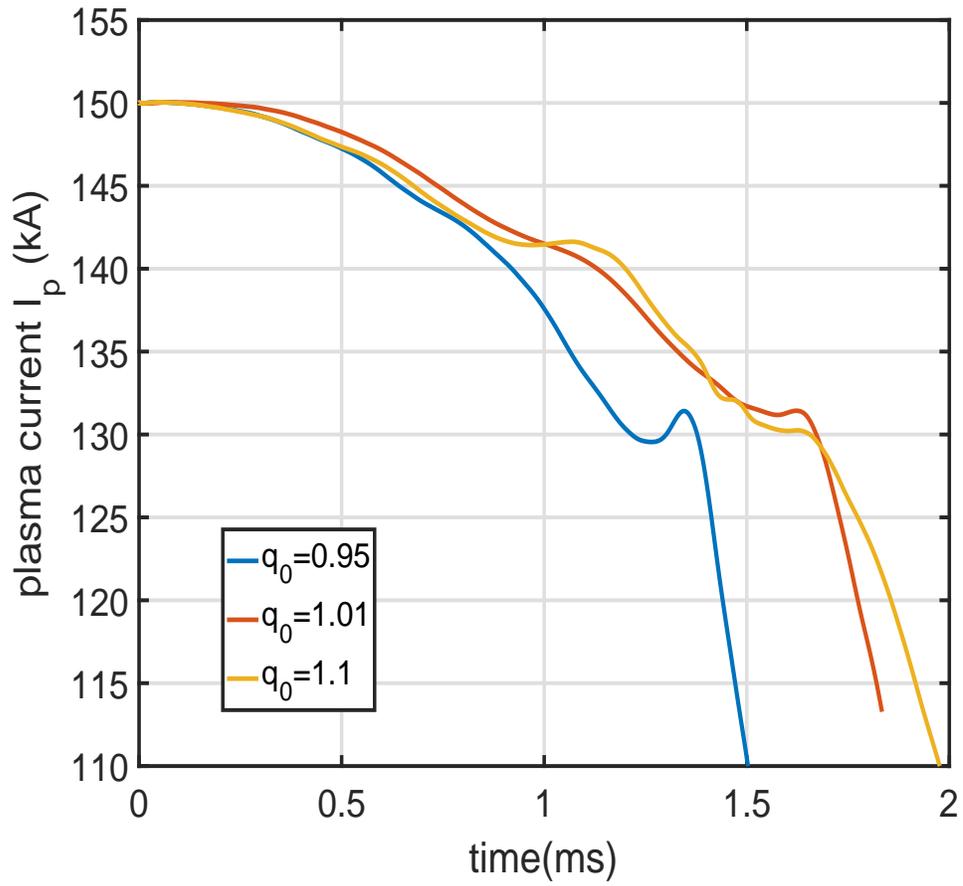}
	\end{center}
	\caption{The plasma current as functions of time for cases with different value of $q_0$ during Argon impurity injections}
	\label{fig:Ar-q0-Ip}
\end{figure}

\begin{figure}[ht]
	\begin{center}
		\includegraphics[width=0.85\textwidth,height=0.5\textheight]{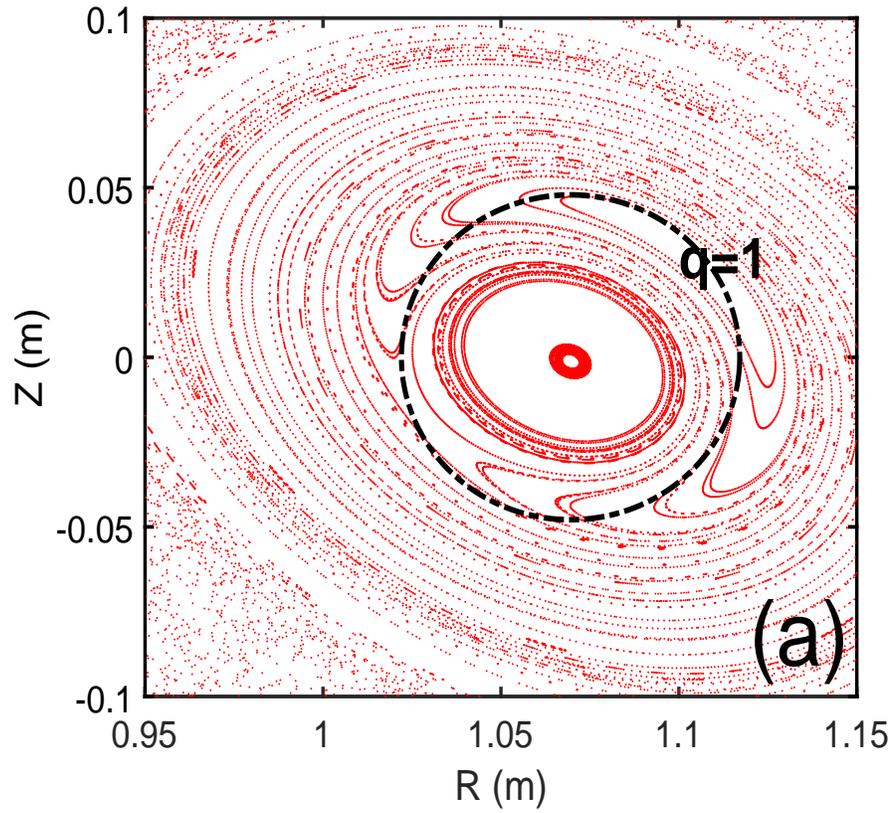}
		\includegraphics[width=0.85\textwidth,height=0.5\textheight]{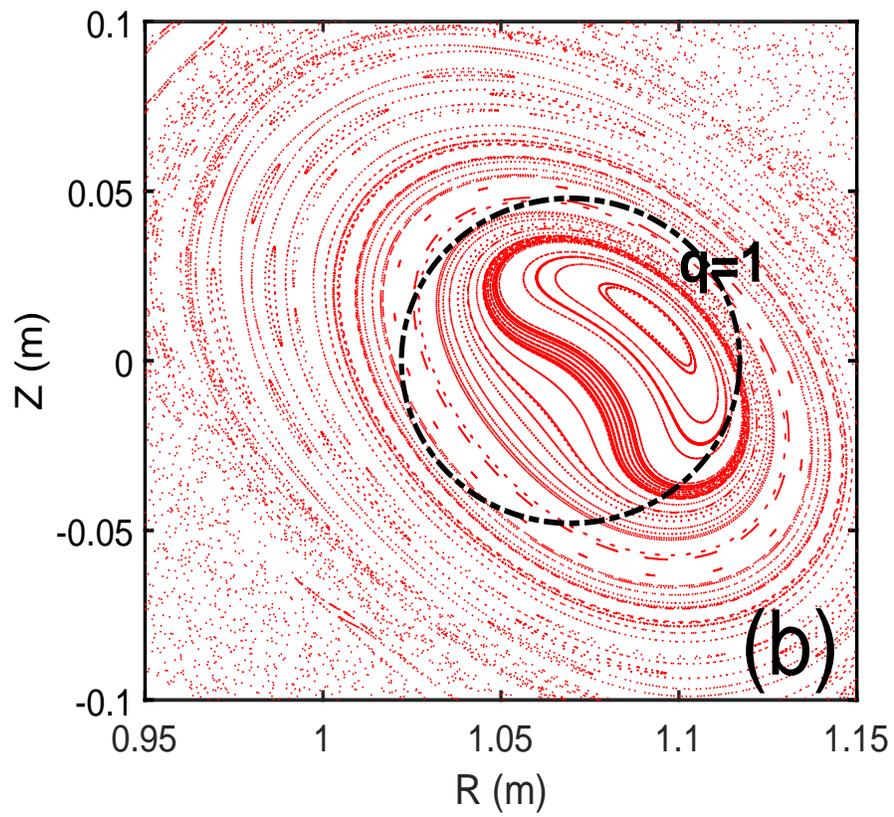}
	\end{center}
	\caption{The Poincare plots in the central region of cases with (a) $q_0=0.95$, and (b) $q_0=1.1$ during Argon impurity injection. The equilibrium $q=1$ surface of the case with $q_0=0.95$ is imposed as black dashed circle for the frame of reference.}
	\label{fig:Ar-q0-poincare}
\end{figure}

In comparison to the Helium impurity used in the above simulations, another set of simulations with the same amount of Argon impurity injection and various values of $q_0$ produce similar results (Fig. \ref{fig:Ar-q0-Ip}). Thus the role of the $q = 1$ surface in the transition between the major disruption and minor disruption appears indifferent to the impurity species, but rather strongly correlated to the presence or absence of reconnection nature in the central MHD activity (Fig. \ref{fig:Ar-q0-poincare}).

\subsection{\label{subsec:imp species}Effect of impurity species}
\begin{figure}[ht]
	\begin{center}
		\includegraphics[width=1.0\textwidth,height=0.35\textheight]{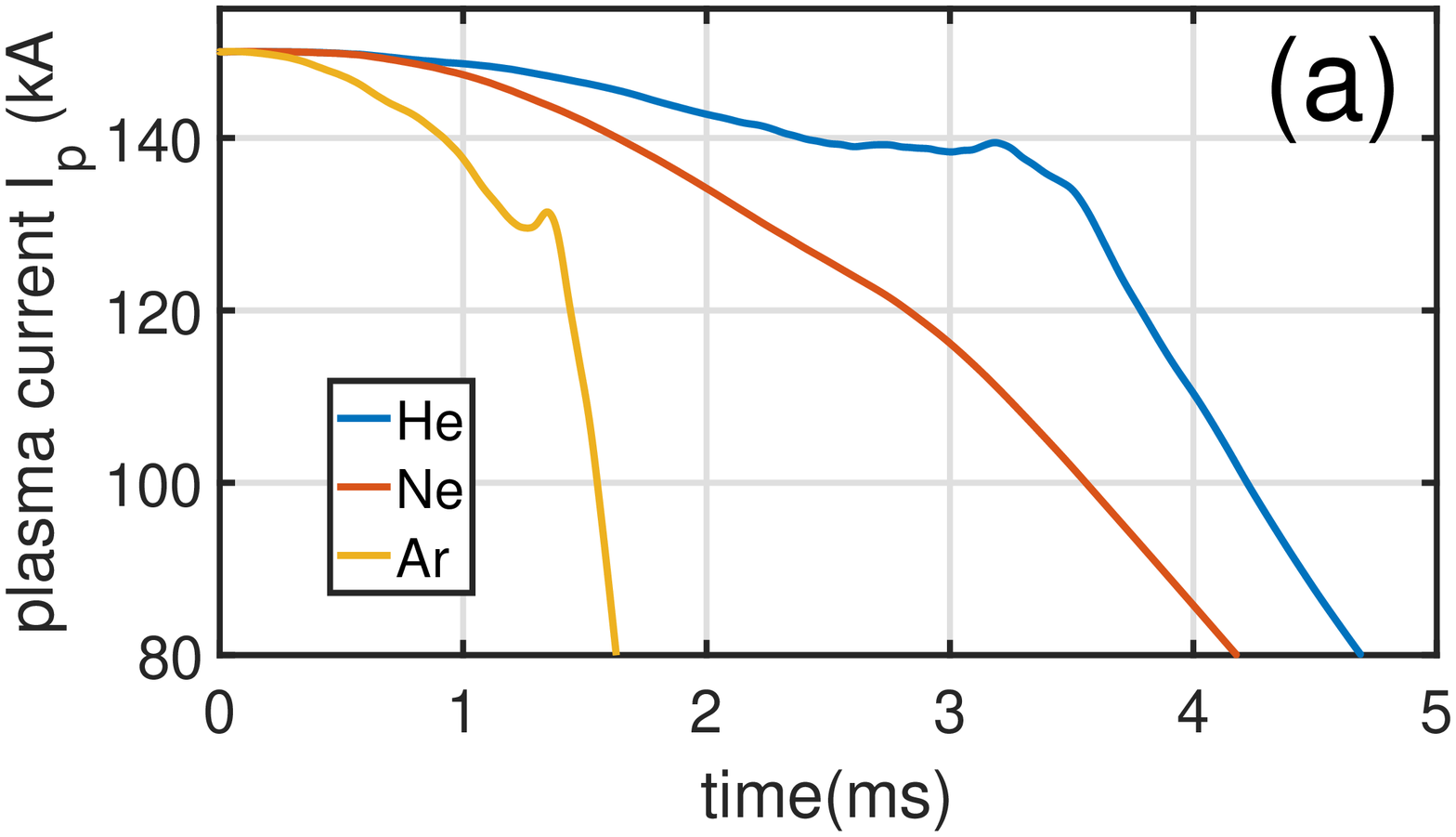}
		\includegraphics[width=1.0\textwidth,height=0.35\textheight]{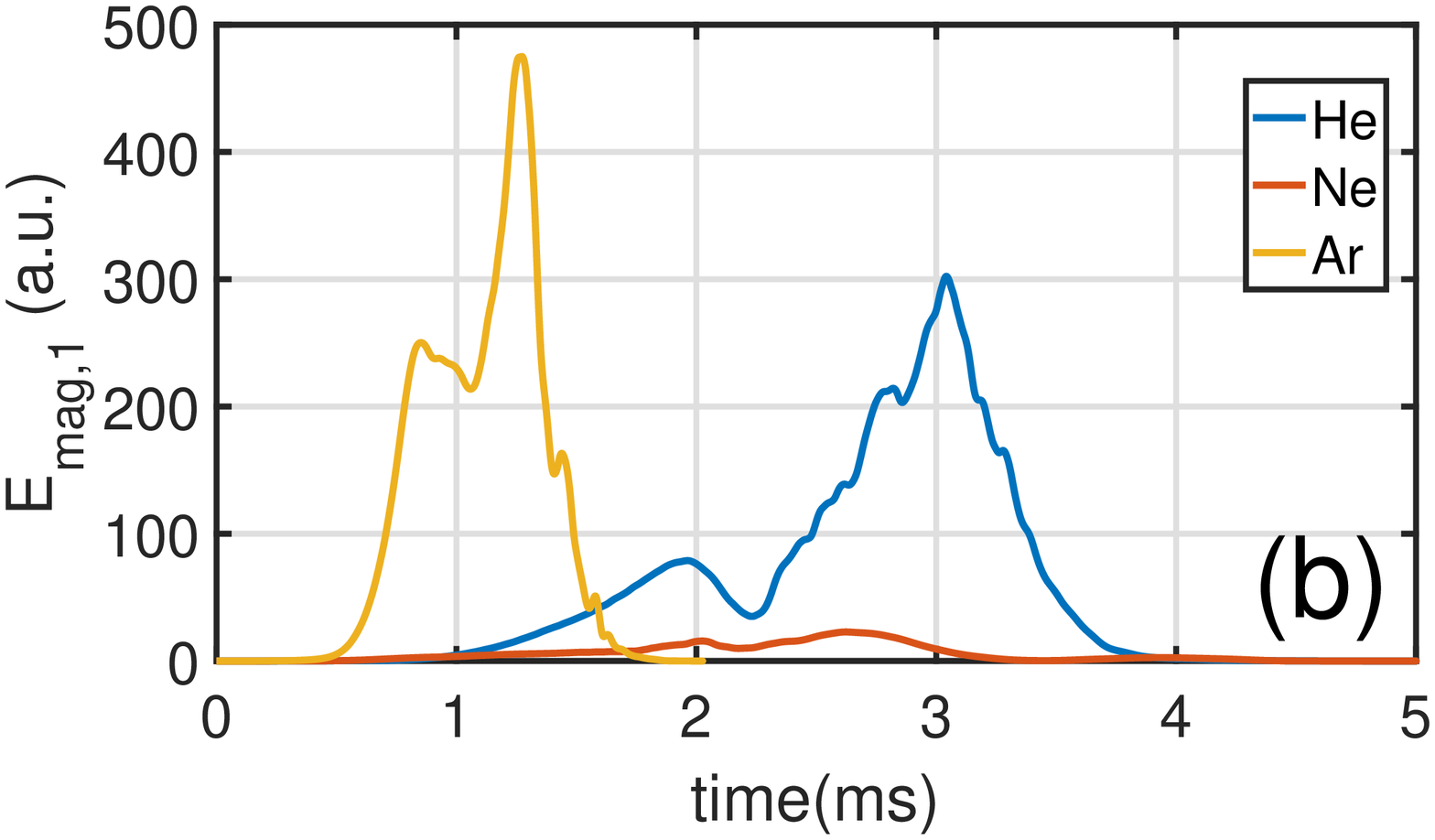}
	\end{center}
	\caption{(a) Plasma current, and (b) perturbed magnetic energies of toroidal component $n=1$ as functions of time for cases with different impurity species.}
	\label{fig:imp-current-n1-peak}
\end{figure}

\begin{figure}[ht]
	\begin{center}
		\includegraphics[width=0.85\textwidth,height=0.5\textheight]{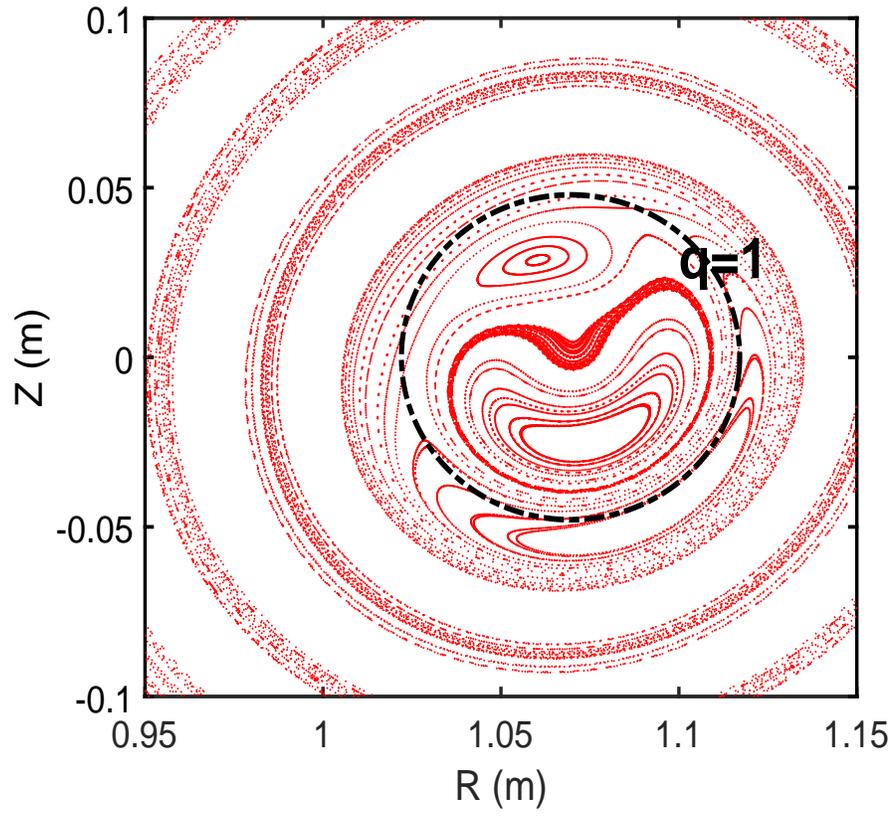}
	\end{center}
	\caption{The Poincare plot in the central region of case during Neon impurity injection.}
	\label{fig:Ne-poincare}
\end{figure}

Simulation results with the same injection level and the safety value at magnetic axis ($q_0 = 0.95$) but different impurity species are compared in Fig. \ref{fig:imp-current-n1-peak}. Naturally, higher Z impurity leads to faster TQ and CQ rates with stronger radiation. The apparent current spike formation is observed in the cases with Argon and Helium injections, but not in the case of Neon injection, and the amplitude of the current spike is approximately proportional to the peak value of the $n=1$ magnetic energy.
A $1/1$ tearing mode is localized in the central region, indicating the occurrence of magnetic reconnection in the cases of Argon and Helium injections, whereas a quasi-interchange-like mode arises in the case with Neon injection (Fig. \ref{fig:Ne-poincare}). This suggests that the impurity species or the ability of radiation can affect the nature and the level of the central MHD activity even for the same equilibrium.

\section{\label{sec:summary}Summary and discussion}
The massive gas injection mitigated disruption process is reproduced using the 3D resistive MHD model implemented in the NIMROD code, which demonstrates the magnetic flux conservation during the pre-TQ phase and its decay in the TQ stage, along with the formation of the current spike at the end of TQ that is correlated to the central MHD activity. The $q = 1$ surface plays a critical role in the transition between the major disruption and minor disruption; namely, when $q_0$ exceeds $1$, the central MHD activity dominated by a local $1/1$ kink-tearing instability is shown to be replaced by a local quasi-interchange like instability, and the TQ stage transits from one major disruption into multiple successive minor disruptions. When $q_0 < 1$, the amplitude of the current spike increases with the decrease of $q_0$, and when $q_0 \ge 1$, the time delay of the final TQ stage increases with the value of $q_0$. In addition, different impurity species has different impacts on the central MHD activity and thus leads to the different behaviors of the current spike even for the same equilibrium and other conditions.

In light of these findings, the potential connection between the current spike, i.e. the large induced loop voltage at the end of TQ phase, and the generation of RE is likely subject to considerable influence from the associated MHD activities. In particular,
the MHD instability and the subsequent stochastization of magnetic field lines during the disruption have been suggested to dissipate the RE effectively in recent experiments \cite{Reux2021}. Thus how the connections among the central MHD activity, the formation of the current spike, and the subsequent TQ process demonstrated in the study reported here may affect the RE dynamics in the ensuing CQ stage is worth further exploration next.

\section{\label{sec:aknoledgments}Acknowledgments}
We are grateful for the supports from the NIMROD team. This work was supported by the National Magnetic Confinement Fusion Program of China (Grant No. 2019YFE03050004), the National Natural Science Foundation of China (Grant Nos. 51821005 and 12175228), Collaborative Innovation Program of Hefei Science Center, CAS (Grant No. 2021HSC-CIP007), and U.S. Department of Energy (Grant Nos. DE-FG02-86ER53218 and DE-SC0018001). This research used the computing resources from the Supercomputing Center of University of Science and Technology of China. 

\section{data availability}
The data that support the findings of this study are available from the corresponding author upon reasonable request.



\newpage

\newpage

\newpage

\newpage

\newpage

\newpage

\newpage

\newpage

\newpage

\newpage


\end{document}